\documentclass[11pt]{article} 
\usepackage[margin=1in]{geometry}
\usepackage{amsmath, amsthm, amsfonts}

\usepackage{graphicx}
\usepackage{subcaption}
\usepackage{url}
\usepackage{booktabs}
\usepackage[table,xcdraw]{xcolor}

\theoremstyle{definition}

\theoremstyle{remark}

\title{India’s Growth Forecast for 2020-21}
\author{Amarendra Das  \\
\small School of Humanities and Social Sciences\\ 
\small amarendra@niser.ac.in \\
\and Subhankar Mishra \\
\small School of Computer Sciences \\
\small smishra@niser.ac.in
\and 
  \small National Institute of Science Education and Research\\
  \small Bhubaneswar-752050, Odisha, India \\
  \small Homi Bhabha National Institute \\
  \small Anushaktinagar, Mumbai - 400094, India\\
}
\date{}
\begin{document}
\maketitle

\abstract{
  COVID-19 has put a severe dent on the global economy and the Indian Economy. International Monetary Fund has projected 1.9 percent for India. However, we believe that due to extended lockdown, the output in the first quarter is almost wiped out. The situation may improve in the second quarter onwards. Nevertheless, due to demand and supply constraints, input constraints, and disruption in the supply chain, except agriculture, no other sector would be able to achieve full capacity of production in 2020-21. The signals from power consumption, GST collection, contraction in the core sectors hint towards a slump in the total output production in 2020-21. We derive the quarterly GVA for 2020-21 based on certain assumptions on the capacity utilisation in different sectors and the quarterly data of 2019-20. We provide quarterly estimates of Gross Value Addition for 2020-21 under two scenarios. We have also estimated the fourth quarter output for 2019-20 under certain assumptions. We estimate a $2.3\%$ growth in 2019-20 and a contraction of output in 2020-21 to the extent of $23$ to $25\%$ . 
}

\section{Impact of COVID-19 on the Economy}
COVID-19 has put a severe dent on the global economy, national economies, and regional economies alike. IMF \cite{IMF:2020} forecasts that the global output is to contract sharply by $-3$ percent in 2020, much worse than during the 2008–09 financial crisis. The advanced economies are set to witness a much higher contraction in the output by $-6.1$ percent and emerging and market economies by $-1$ percent. It also further warns that the risks for even more severe outcomes are substantial. For India, different agencies have provided different growth estimates for 2020-21. International Monetary Fund has projected 1.9 percent; Fitch Ratings \cite{ET1:2020} has forecast $0.8\%$ growth. In early April ADB had forecast a $4\%$ for India during 2020-21. Professor C. Rangarajan and D. K Srivastav \cite{TheHindu:2020}  have estimated $4.4\%$ growth for India in 2019-20 and $2.94\%$ in 2020-21. The professional forecasters' projection of real GDP growth given in the monetary policy report of Reserve Bank of India released in April 2020 reports $4.6\%$ growth in the last quarter of 2019-20, $4.7\%$ in first quarter, $5.3\%$ in second quarter, $5.7\%$ in third quarter and $6.1\%$ in fourth quarter. However, given the length of complete lockdown in India and the suspension of most of the economic activities in the first quarter it is hard to believe that the Indian economy will achieve any growth in 2020-21. We apprehend that the total output in India will contract drastically in the first quarter, although there will be some improvement in the subsequent quarters, growth of total output won’t be higher than the last year. Therefore, the fear of a double-digit contraction in the output looms large. 

\begin{table}[]
  \centering
  \begin{tabular}{@{}lr@{}} 
  \toprule
  \textbf{Agency}                            & \textbf{Growth Prediction} \\ \midrule
  IMF  \cite{IMF:2020}                                      & 1.9\%                      \\
  Fitch Ratings \cite{ET1:2020}                                    & 0.8\%                      \\
  ADB  \cite{ET4:2020}                                      & 4\%                        \\
  Professor C. Rangarajan and D. K Srivastav  \cite{TheHindu:2020}& 4.4\%                      \\
  \textbf{Our Prediction}                            & \textbf{-23\%}                      \\ \bottomrule
  \end{tabular}%
  \caption{Growth Prediction for India FY 2020-21}
  \label{tab:my-table}
  \end{table}

The nationwide lockdown was imposed on March 25. This suspended almost all economic activities including the operation of factories, construction activities, running of trains, buses, and flights. There are many signals of the fall in the output in the first quarter. The power consumption in April 2020 has declined by 22.75\% as compared to April 2019 \cite{ET2:2020}. The core sector output fell by 6.5\% in March 2020 with just one week of lockdown in the country \cite{ET3:2020}. The Gross Direct Tax collection in April had declined by 5.4\% \cite{CMIE:2020}. The tax collection by the Income Tax Department in April is the tax collected by the employers for the income of March. Therefore, the income tax collection for the month of April will be reflected in May. This may show a significant decline in direct tax collection. This is vindicated from the fact that most of the private companies have slashed the salary of employees up to 50\%. Even in the public sector, the Union government froze the Dearness Allowances for one year. Many state governments have slashed the salary of government employees. The GST collection of states has declined by more than 75\% \cite{BusinessToday:2020}. The Union government has deferred the release of the GST collection data for April month. The CMIE data further shows that the unemployment rate in the first week of May has gone up to 23.8 percent. Most of the metros in India are coming under the red zones, where more than one-third of the total output is produced. Due to the movement of migrant workers, the number of districts coming under red zones are increasing every day. In this context it would be unwise to expect any normal operation of economic activities soon. All these signals a drastic fall in the total output and income in the country during the first quarter of 2020-21. Therefore, we believe that instead of any growth the Indian economy will witness a slump in the total output.

If we assume that the lockdown will be relaxed from mid-may, then one and half month of the first quarter is going under lockdown. Although the state and national governments have relaxed some selected economic activities in lockdown 3.0 starting from May 04, it is far from the normal operation of those sectors. Even if the lockdown is relaxed from mid may, there will be a lot of restrictions in most of the economic activities. Therefore, the effect of COVID may continue in the second quarter and third quarters. We may expect normal activities in the fourth quarter. If we look at each sector independently during this fiscal year, agriculture is less affected compared to all other sectors. The manufacturing and service sectors are the worst hit sectors due to lockdown. Due to COVID, this export would be badly affected. Due to massive unemployment, and the return of migrant workers from within the country and outside, the purchasing power of the people will shrink. Thus the aggregate demand (internal + external) will decline significantly. 

In this context we have attempted to calculate the GSVA of India for the last quarter of 2019-20 and all four quarters of 2020-21.

\section{Methodology} 
We have used the GVA data provided by RBI in its website database on the Indian economy. We have used the quarterly GVA data for 2018-19, 2019-20 in 2011-12 constant prices. The GVA data for the fourth quarter of 2019-20 are not available. Therefore, first of all we estimate the same using two broad assumptions. Since the lockdown was imposed in the last week of the last quarter it would have impacted the GVA in all sectors badly. In March, all government and private agencies try to achieve the targets. Even though the nationwide lockdown was imposed from March 25, many states had already imposed a lot of restrictions from mid-March. Keeping this scenario in mind we assume that only the agriculture sector would not witness any contraction and produce as much as it had produced last year. All other sectors would witness a 6 percent contraction. This assumption is based on the findings that in March the core sectors witnessed 6 percent contraction. Using the fourth quarter data of 2018-19 we estimated the fourth quarter GVA under these assumptions.  Aggregating all the quarterly data for 2019-20 we derived the GVA for 2019-20 and estimated the quarterly growth rate and annual growth rates for the year 2019-20. 

Next we calculate the GVA for the year 2020-21. At the outset we assume that in the absence of any growth, the nation has the capacity to produce at least the same level of output that was produced in 2019-20. But the real output would be much different. Due to lockdown in the first quarter, the output of one and half quarters is almost wiped out, barring a few sectors like agriculture, banking, public administration, and defense, etc.  The entire country could have used around 25\% of the output capacity in the first quarter. This assumption can be supported by different other indicators like reduction in power consumption, GST collection, Therefore, we assume that, the in most of the sectors, full output potential cannot be realized at least in the first two quarters. In the second quarter even though the lockdown would not be there, life will not come to complete normalcy. Moreover, many sectors would be facing the demand constraints, supply chain disruption, and input supply problem. Most of the industries would face labour scarcity, as migrant workers are coming back to their native places. Some sectors may achieve normalcy in the third quarter due to harvest of Kharif and rise in purchase power. The fourth quarter may see some growth. We have assumed that wherever (sectors) there is scope, the fourth quarter may witness growth that of the previous year.  Table \ref{tab:S1Util202021} and Table \ref{tab:S2Util202021} show the assumptions on the capacity utilization in different sectors in four quarters of 2020-21 under scenario 1 and scenario 2. Scenario 2 is a more pessimistic scenario. In this we assume that even the third quarter of 2020-21 may not be fully normal and the fourth quarter may not see any growth in most of the sectors.

\section{Results}
\begin{figure}[ht]
  \centering
  \includegraphics[width=0.5\textwidth]{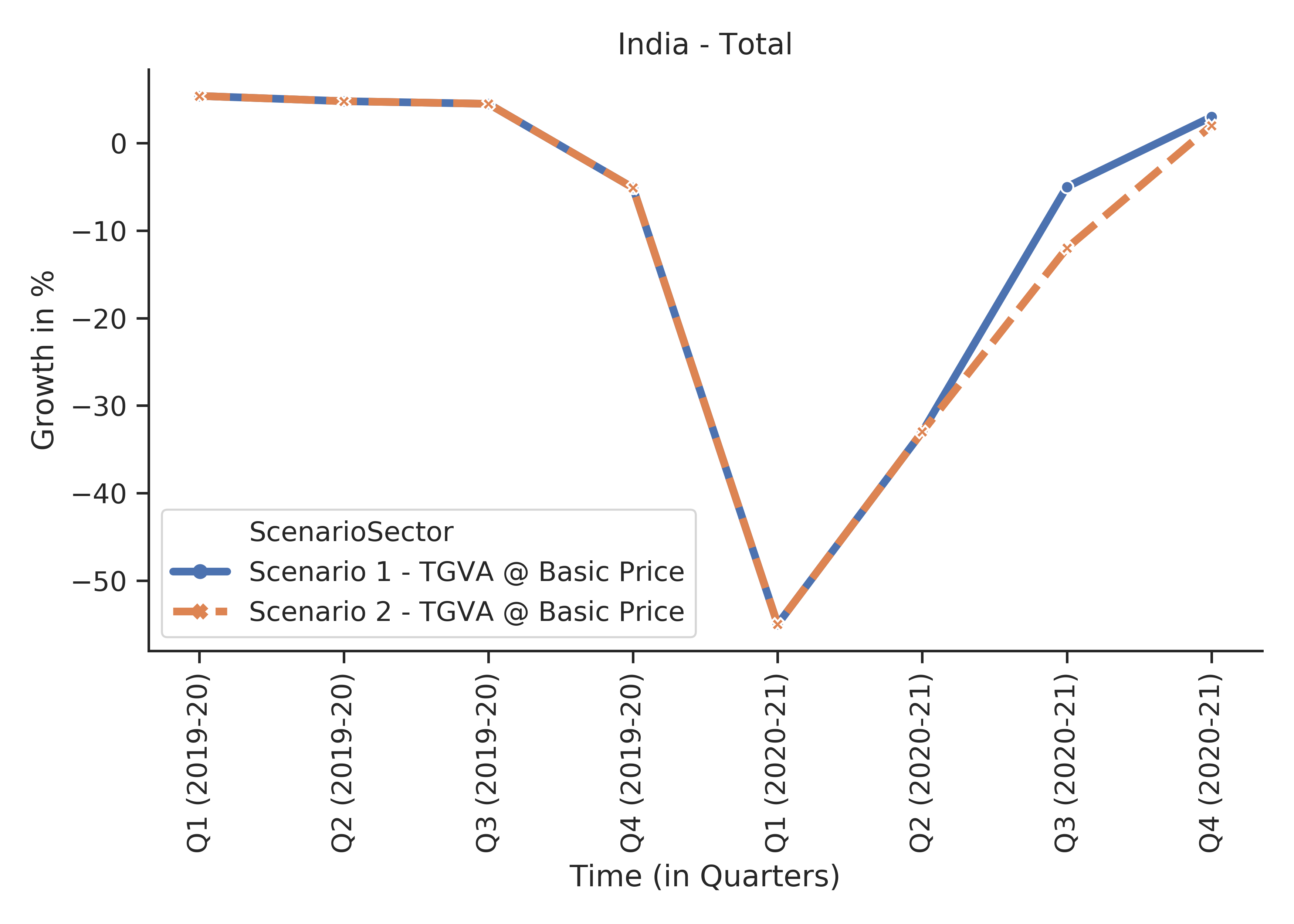}
  \caption{Quarterly Growth of GVA during 2019-20 and 2020-21 under Scenario 1 and Scenario}
  \label{fig:g}
\end{figure}

\begin{figure}[ht]
  \begin{subfigure}{.45\textwidth}
    \centering
    \includegraphics[width=.8\linewidth]{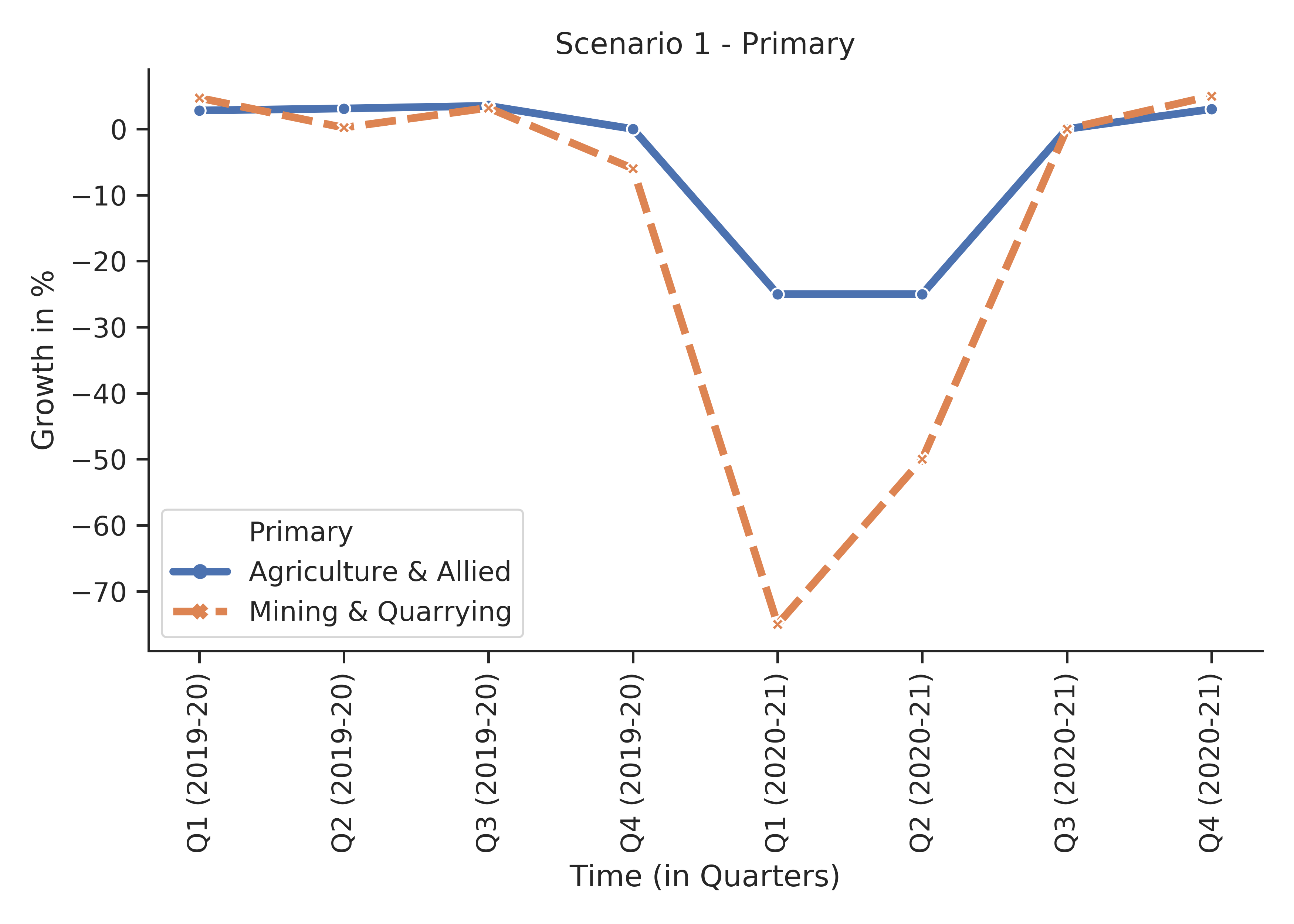}  
    \caption{Scenario 1}
  \end{subfigure}
  \begin{subfigure}{.45\textwidth}
    \centering
    \includegraphics[width=.8\linewidth]{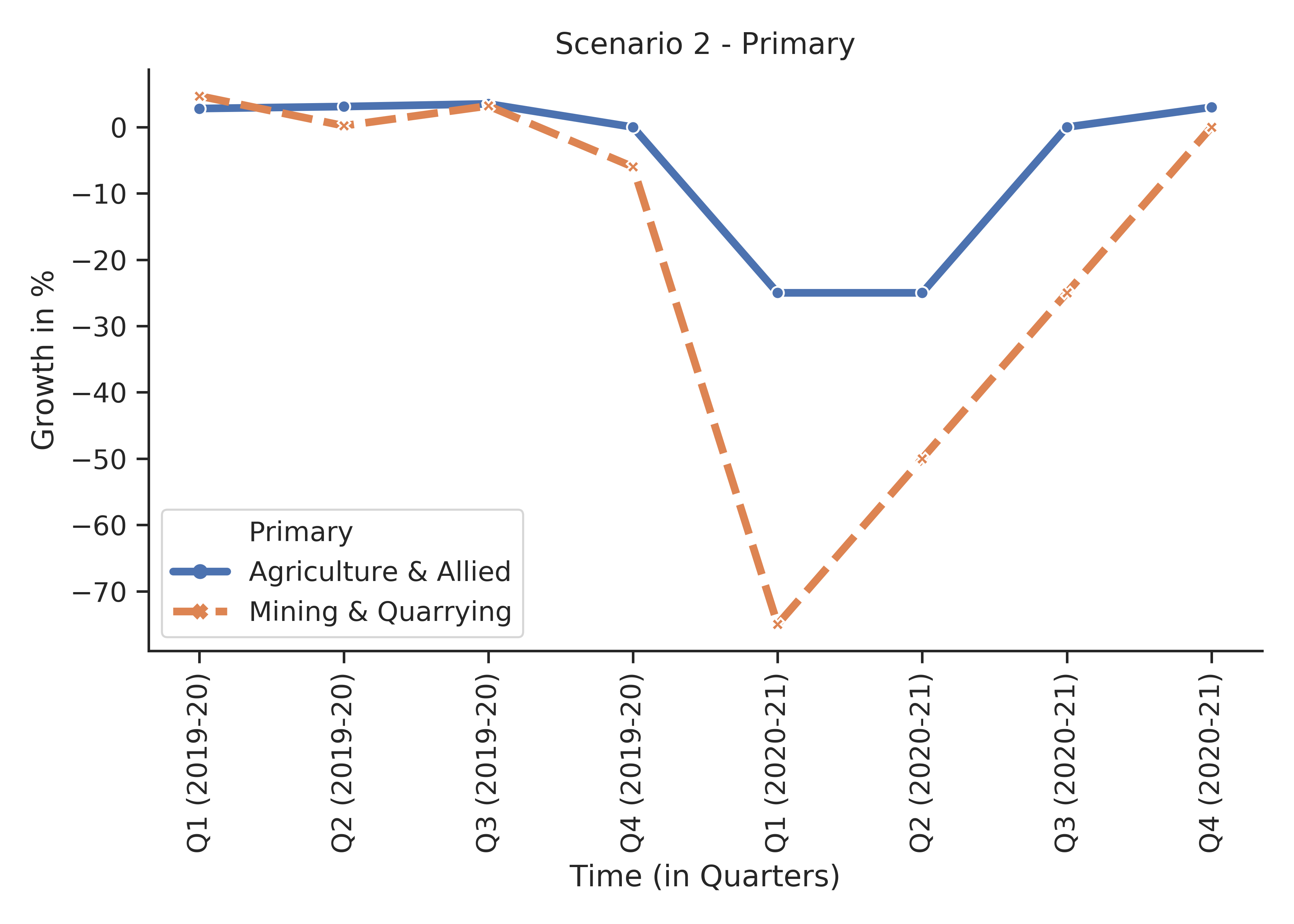}  
    \caption{Scenario 2}
  \end{subfigure}
  \caption{Quarterly Growth of GVA in Primary Sector during 2019-20 and 2020-21}
  \label{fig:pg}
  \end{figure}

\begin{figure}[ht]
    \begin{subfigure}{.45\textwidth}
      \centering
      \includegraphics[width=.8\linewidth]{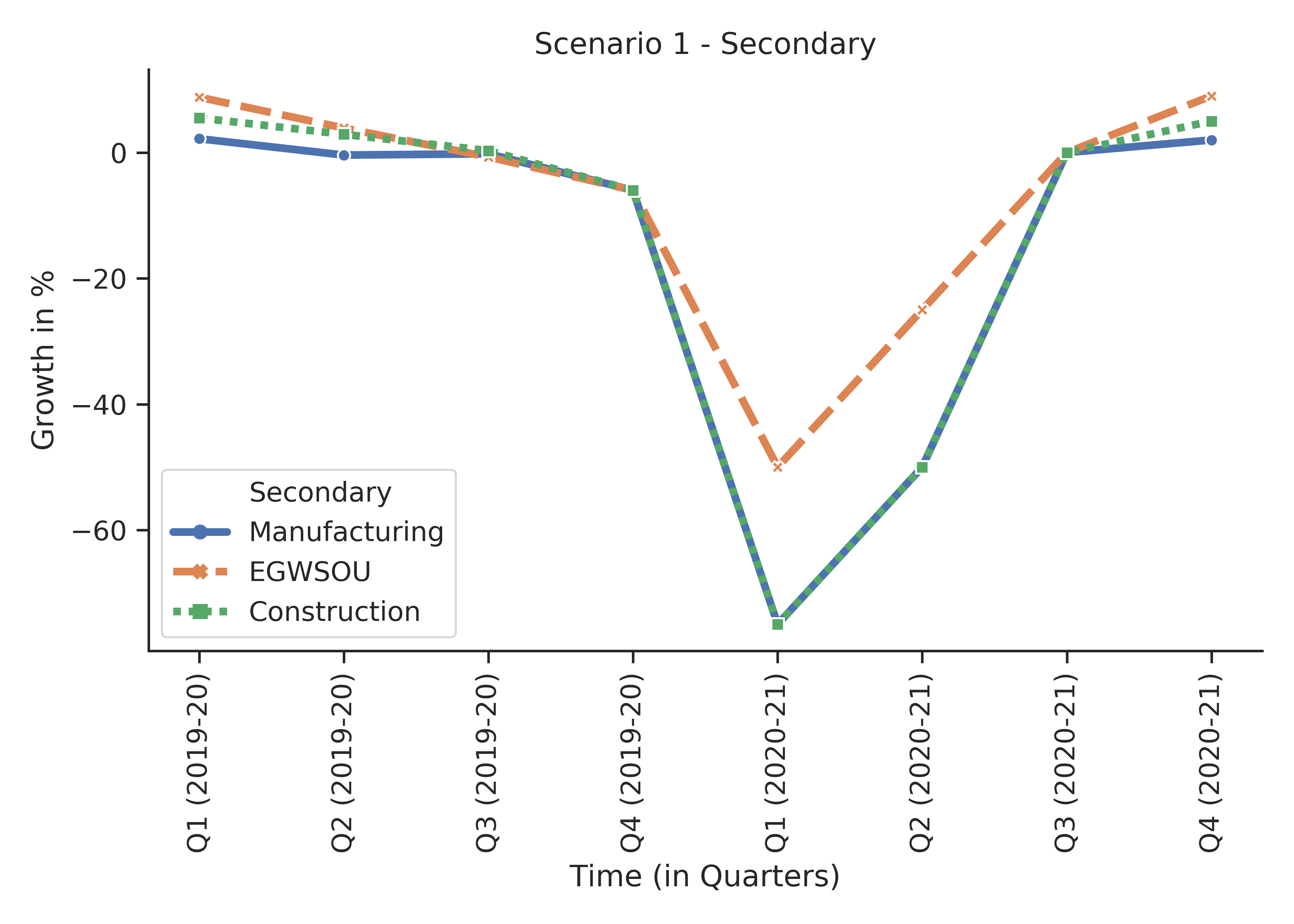}  
      \caption{Scenario 1}
    \end{subfigure}
    \begin{subfigure}{.45\textwidth}
      \centering
      \includegraphics[width=.8\linewidth]{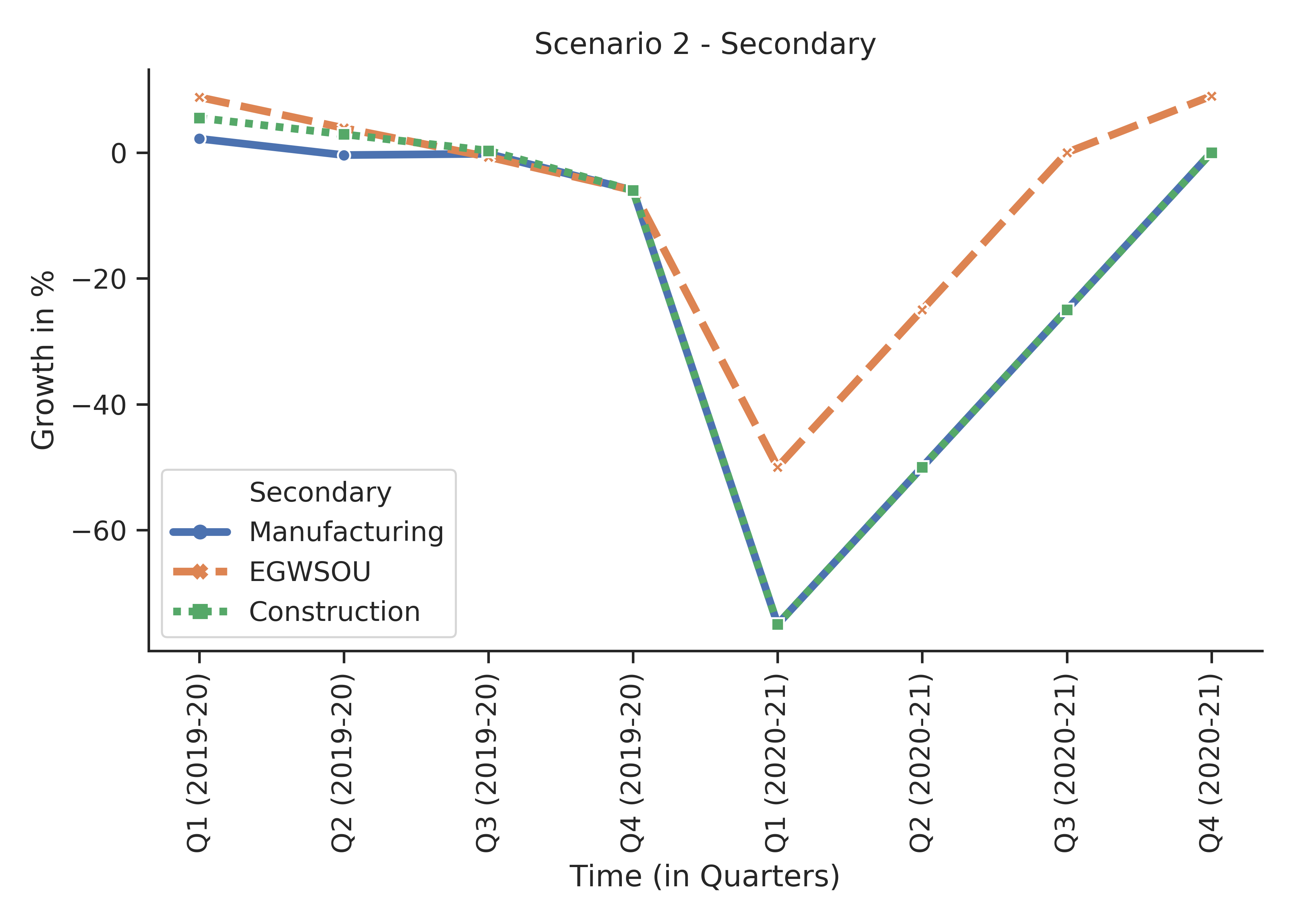}  
      \caption{Scenario 2}
    \end{subfigure}
    \caption{Quarterly Growth of GVA in Secondary Sector during 2019-20 and 2020-21}
    \label{fig:sg}
    \end{figure}

\begin{figure}[ht]
      \begin{subfigure}{.45\textwidth}
        \centering
        \includegraphics[width=.8\linewidth]{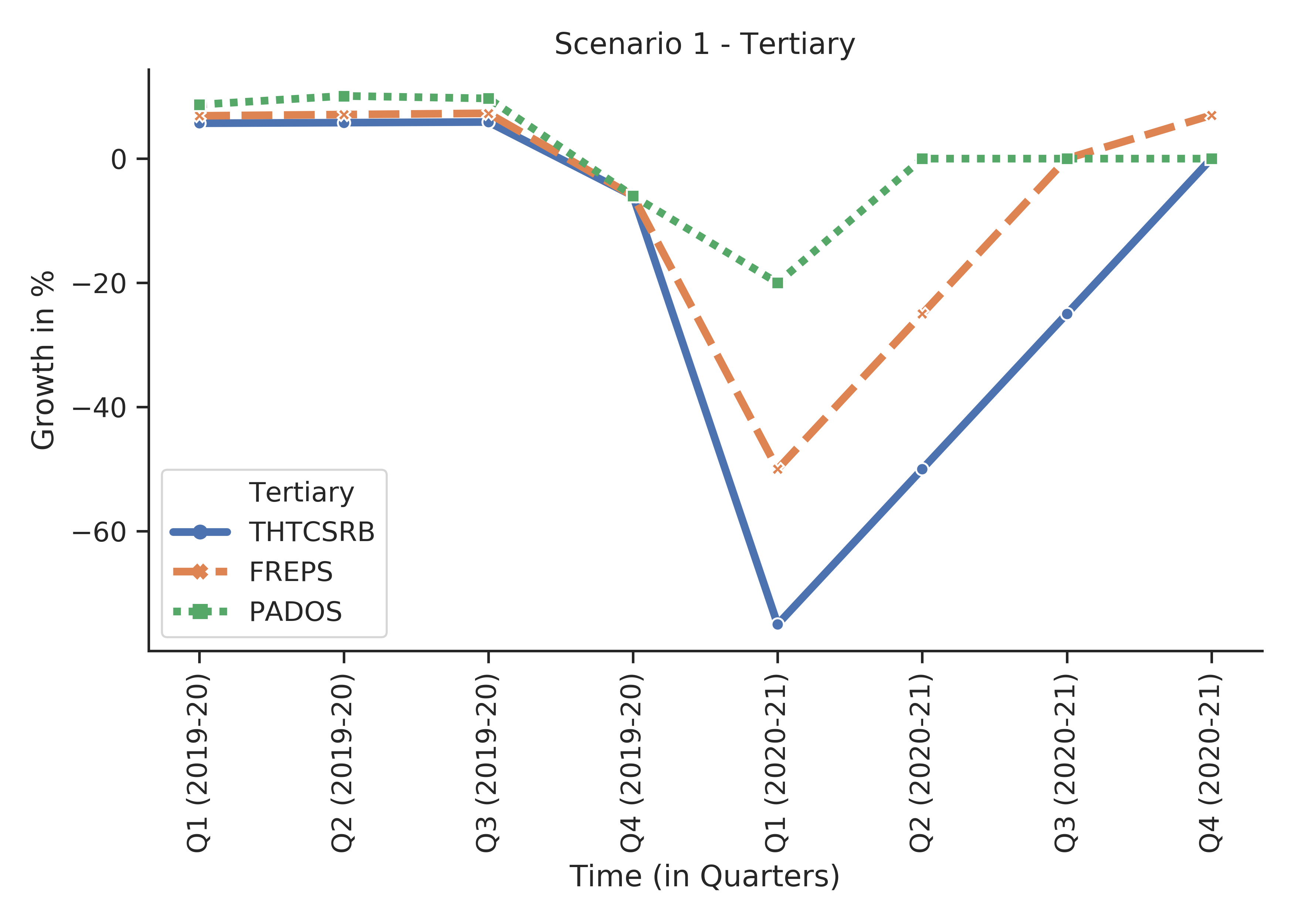}  
        \caption{Scenario 1}
      \end{subfigure}
      \begin{subfigure}{.45\textwidth}
        \centering
        \includegraphics[width=.8\linewidth]{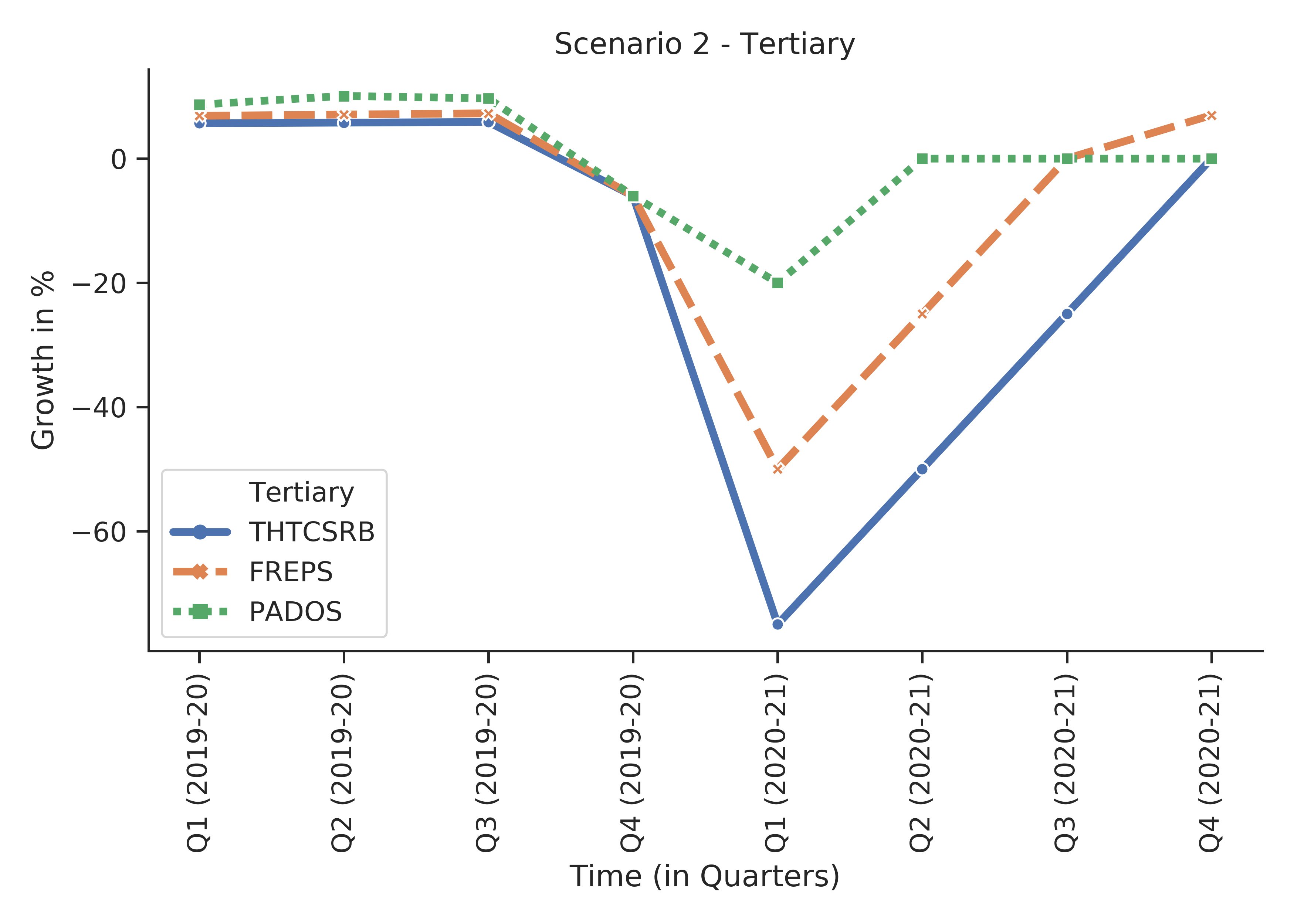}  
        \caption{Scenario 2}
      \end{subfigure}
      \caption{Quarterly Growth of GVA in Tertiary Sector during 2019-20 and 2020-21}
      \label{fig:tg}
      \end{figure}

Figure \ref{fig:g} shows the quarterly growth rates of GVA under scenario 1 and 2.  Figure \ref{fig:pg}, \ref{fig:sg} and  \ref{fig:tg} show the quarterly growth of primary, secondary and tertiary sectors under scenario 1 and scenario 2. Table \ref{tab:QGVA201920} shows the GVA estimates for 2019-20 and Table \ref{tab:QG201920} shows the quarterly growth rates for the year 2019-20. We estimate a 2.3\% real growth in 2019-20. Table \ref{tab:S1QGVA202021} presents the quarterly estimates of GVA in 2020-21 under scenario 1 and Table \ref{tab:S1QG202021} presents the quarterly growth rates of GVA in various sectors and total GVA under Scenario 1. We estimate a 23 percent contraction in the GVA in 2020-21. Table \ref{tab:S2QGVA202021} presents the quarterly estimates of GVA for 2020-21 under scenario 2 and Table \ref{tab:S2QG202021} presents the quarterly growth rates of GVA under scenario 2.  Under scenario 2 we estimate a 25 percent contraction of GVA.  

\section{Conclusion}
The prolonged lock down in India has a huge economic cost.    Although different agencies have estimated a positive growth for India in 2020-21, we   differ with this. Our estimates suggest that Indian economy is set to witness a major slump in 2020-21 to the tune of 23 to 25 percent. This is due to nearly complete suspension of all economic activities in the first quarter and possible disruption of productions in the subsequent quarters owing demand and supply bottlenecks, disruptions in the input supply, output supply chain and labour shortages.

\begin{table}[]
  \centering
  \caption{Quarterly Gross Value Addition in 2019-20 in Lakh Crore INR. Q4 has been estimated by assuming a no growth in agriculture and hence the output of 4th quarter data of 2018-19 has been used. For other sectors We have taken a 6\% fall in the output based on the findings on the 6\% fall in the core sector output.}
  \label{tab:QGVA201920}
  \resizebox{\textwidth}{!}{%
  \begin{tabular}{@{}lrrrrr@{}}
  \toprule
  {\color[HTML]{000000} Quarter} &
    {\color[HTML]{000000} \textbf{Q1}} &
    {\color[HTML]{000000} \textbf{Q2}} &
    {\color[HTML]{000000} \textbf{Q3}} &
    {\color[HTML]{000000} \textbf{Q4}} &
    {\color[HTML]{000000} \textbf{Total}} \\ \midrule
  \rowcolor[HTML]{F2F2F2} 
  {\color[HTML]{000000} Agriculture, Forestry and Fishing} &
    {\color[HTML]{000000} 439248} &
    {\color[HTML]{000000} 366429} &
    {\color[HTML]{000000} 609105} &
    {\color[HTML]{000000} 486094} &
    {\color[HTML]{000000} \textbf{1900876}} \\
  {\color[HTML]{000000} Mining \& Quarrying} &
    {\color[HTML]{000000} 92777} &
    {\color[HTML]{000000} 69890} &
    {\color[HTML]{000000} 85429} &
    {\color[HTML]{000000} 107147} &
    {\color[HTML]{000000} \textbf{355243}} \\
  \rowcolor[HTML]{F2F2F2} 
  {\color[HTML]{000000} Manufacturing} &
    {\color[HTML]{000000} 574411} &
    {\color[HTML]{000000} 577184} &
    {\color[HTML]{000000} 559335} &
    {\color[HTML]{000000} 579744} &
    {\color[HTML]{000000} \textbf{2290674}} \\
  {\color[HTML]{000000} Electricity, Gas, Water Supply \& Other Utility} &
    {\color[HTML]{000000} 81628} &
    {\color[HTML]{000000} 79525} &
    {\color[HTML]{000000} 72817} &
    {\color[HTML]{000000} 65052} &
    {\color[HTML]{000000} \textbf{299022}} \\
  \rowcolor[HTML]{F2F2F2} 
  {\color[HTML]{000000} Construction} &
    {\color[HTML]{000000} 263653} &
    {\color[HTML]{000000} 244863} &
    {\color[HTML]{000000} 260170} &
    {\color[HTML]{000000} 243097} &
    {\color[HTML]{000000} \textbf{1011783}} \\
  {\color[HTML]{000000} Trade, Hotels, Transport, Communication and Broadcasting} &
    {\color[HTML]{000000} 644224} &
    {\color[HTML]{000000} 621153} &
    {\color[HTML]{000000} 645479} &
    {\color[HTML]{000000} 635517} &
    {\color[HTML]{000000} \textbf{2546373}} \\
  \rowcolor[HTML]{F2F2F2} 
  {\color[HTML]{000000} Financial, Real Estate and Professional Services} &
    {\color[HTML]{000000} 810500} &
    {\color[HTML]{000000} 873813} &
    {\color[HTML]{000000} 643389} &
    {\color[HTML]{000000} 587109} &
    {\color[HTML]{000000} \textbf{2914811}} \\
  {\color[HTML]{000000} Public Administration, Defence and Other Services} &
    {\color[HTML]{000000} 421191} &
    {\color[HTML]{000000} 461487} &
    {\color[HTML]{000000} 474947} &
    {\color[HTML]{000000} 433359} &
    {\color[HTML]{000000} \textbf{1790984}} \\
  \rowcolor[HTML]{F2F2F2} 
  {\color[HTML]{000000} Total Gross Value Added at Basic Price} &
    {\color[HTML]{000000} 3327632} &
    {\color[HTML]{000000} 3294343} &
    {\color[HTML]{000000} 3350669} &
    {\color[HTML]{000000} 3137119} &
    {\color[HTML]{000000} \textbf{13109763}} \\ \bottomrule
  \end{tabular}
  }
  \end{table}

\begin{table}[]
  \centering
  \caption{Quarterly and Annual Growth Rate in 2019-20. Source (Author's Estimation)}
  \label{tab:QG201920}
  \resizebox{\textwidth}{!}{%
  \begin{tabular}{@{}lrrrrr@{}}
    \toprule
    {\color[HTML]{000000} Quarter} &
      {\color[HTML]{000000} \textbf{Q1}} &
      {\color[HTML]{000000} \textbf{Q2}} &
      {\color[HTML]{000000} \textbf{Q3}} &
      {\color[HTML]{000000} \textbf{Q4}} &
      {\color[HTML]{000000} \textbf{Total}} \\ \midrule
    \rowcolor[HTML]{F2F2F2} 
    {\color[HTML]{000000} Agriculture, Forestry and Fishing} &
      {\color[HTML]{000000} 2.8\%} &
      {\color[HTML]{000000} 3.1\%} &
      {\color[HTML]{000000} 3.5\%} &
      {\color[HTML]{000000} 0.0\%} &
      {\color[HTML]{000000} 2.3\%} \\
    {\color[HTML]{000000} Mining \& Quarrying} &
      {\color[HTML]{000000} 4.7\%} &
      {\color[HTML]{000000} 0.2\%} &
      {\color[HTML]{000000} 3.2\%} &
      {\color[HTML]{000000} -6.0\%} &
      {\color[HTML]{000000} 0.0\%} \\
    \rowcolor[HTML]{F2F2F2} 
    {\color[HTML]{000000} Manufacturing} &
      {\color[HTML]{000000} 2.2\%} &
      {\color[HTML]{000000} -0.4\%} &
      {\color[HTML]{000000} -0.2\%} &
      {\color[HTML]{000000} -6.0\%} &
      {\color[HTML]{000000} -1.2\%} \\
    {\color[HTML]{000000} Electricity, Gas, Water Supply \& Other Utility} &
      {\color[HTML]{000000} 8.8\%} &
      {\color[HTML]{000000} 3.9\%} &
      {\color[HTML]{000000} -0.7\%} &
      {\color[HTML]{000000} -6.0\%} &
      {\color[HTML]{000000} 1.7\%} \\
    \rowcolor[HTML]{F2F2F2} 
    {\color[HTML]{000000} Construction} &
      {\color[HTML]{000000} 5.5\%} &
      {\color[HTML]{000000} 2.9\%} &
      {\color[HTML]{000000} 0.3\%} &
      {\color[HTML]{000000} -6.0\%} &
      {\color[HTML]{000000} 0.6\%} \\
    {\color[HTML]{000000} Trade, Hotels, Transport, Communication and Broadcasting} &
      {\color[HTML]{000000} 5.7\%} &
      {\color[HTML]{000000} 5.8\%} &
      {\color[HTML]{000000} 5.9\%} &
      {\color[HTML]{000000} -6.0\%} &
      {\color[HTML]{000000} 2.6\%} \\
    \rowcolor[HTML]{F2F2F2} 
    {\color[HTML]{000000} Financial, Real Estate and Professional Services} &
      {\color[HTML]{000000} 6.9\%} &
      {\color[HTML]{000000} 7.1\%} &
      {\color[HTML]{000000} 7.3\%} &
      {\color[HTML]{000000} -6.0\%} &
      {\color[HTML]{000000} 4.2\%} \\
    {\color[HTML]{000000} Public Administration, Defence and Other Services} &
      {\color[HTML]{000000} 8.7\%} &
      {\color[HTML]{000000} 10.1\%} &
      {\color[HTML]{000000} 9.7\%} &
      {\color[HTML]{000000} -6.0\%} &
      {\color[HTML]{000000} 5.3\%} \\
    \rowcolor[HTML]{F2F2F2} 
    {\color[HTML]{000000} Total Gross Value Added at Basic Price} &
      {\color[HTML]{000000} 5.4\%} &
      {\color[HTML]{000000} 4.8\%} &
      {\color[HTML]{000000} 4.5\%} &
      {\color[HTML]{000000} -5.1\%} &
      {\color[HTML]{000000} 2.3\%} \\ \bottomrule
    \end{tabular}
  }
  \end{table}

  \begin{table}[]
    \centering
    \caption{Scenario 1 - Assumptions on the Capacity Utilisation in different major sectors four quarters during 2020-21 (Author's Estimation)}
    \label{tab:S1Util202021}
    \begin{tabular}{@{}lrrrrr@{}}
    \toprule
    {\color[HTML]{000000} Quarter} &
      {\color[HTML]{000000} \textbf{Q1}} &
      {\color[HTML]{000000} \textbf{Q2}} &
      {\color[HTML]{000000} \textbf{Q3}} &
      {\color[HTML]{000000} \textbf{Q4}} \\ \midrule
    \rowcolor[HTML]{F2F2F2} 
    {\color[HTML]{000000} Agriculture, Forestry and Fishing} &
      {\color[HTML]{000000} 0.75} &
      {\color[HTML]{000000} 0.75} &
      {\color[HTML]{000000} 1.00} &
      {\color[HTML]{000000} 1.03} \\
    {\color[HTML]{000000} Mining \& Quarrying} &
      {\color[HTML]{000000} 0.25} &
      {\color[HTML]{000000} 0.50} &
      {\color[HTML]{000000} 1.00} &
      {\color[HTML]{000000} 1.05} \\
    \rowcolor[HTML]{F2F2F2} 
    {\color[HTML]{000000} Manufacturing} &
      {\color[HTML]{000000} 0.25} &
      {\color[HTML]{000000} 0.50} &
      {\color[HTML]{000000} 1.00} &
      {\color[HTML]{000000} 1.02} \\
    {\color[HTML]{000000} Electricity, Gas, Water Supply \& Other Utility} &
      {\color[HTML]{000000} 0.50} &
      {\color[HTML]{000000} 0.75} &
      {\color[HTML]{000000} 1.00} &
      {\color[HTML]{000000} 1.09} \\
    \rowcolor[HTML]{F2F2F2} 
    {\color[HTML]{000000} Construction} &
      {\color[HTML]{000000} 0.25} &
      {\color[HTML]{000000} 0.50} &
      {\color[HTML]{000000} 1.00} &
      {\color[HTML]{000000} 1.06} \\
    {\color[HTML]{000000} Trade, Hotels, Transport, Communication and Broadcasting} &
      {\color[HTML]{000000} 0.25} &
      {\color[HTML]{000000} 0.50} &
      {\color[HTML]{000000} 0.75} &
      {\color[HTML]{000000} 1.00} \\
    \rowcolor[HTML]{F2F2F2} 
    {\color[HTML]{000000} Financial, Real Estate and Professional Services} &
      {\color[HTML]{000000} 0.50} &
      {\color[HTML]{000000} 0.75} &
      {\color[HTML]{000000} 1.00} &
      {\color[HTML]{000000} 1.07} \\
    {\color[HTML]{000000} Public Administration, Defence and Other Services} &
      {\color[HTML]{000000} 0.80} &
      {\color[HTML]{000000} 1.00} &
      {\color[HTML]{000000} 1.00} &
      {\color[HTML]{000000} 1.00} \\ \bottomrule
    \end{tabular}
  \end{table}

  \begin{table}[]
    \centering
    \caption{Scenario 2 - Assumptions on the Capacity Utilisation in different major sectors four quarters during 2020-21 (Author's Estimation)}
    \label{tab:S2Util202021}
    \begin{tabular}{@{}lrrrrr@{}}
      \toprule
    {\color[HTML]{000000} Quarter} &
      {\color[HTML]{000000} \textbf{Q1}} &
      {\color[HTML]{000000} \textbf{Q2}} &
      {\color[HTML]{000000} \textbf{Q3}} &
      {\color[HTML]{000000} \textbf{Q4}} \\ \midrule
    \rowcolor[HTML]{F2F2F2} 
    {\color[HTML]{000000} Agriculture, Forestry and Fishing} &
      {\color[HTML]{000000} 0.75} &
      {\color[HTML]{000000} 0.75} &
      {\color[HTML]{000000} 1.00} &
      {\color[HTML]{000000} 1.03} \\
    {\color[HTML]{000000} Mining \& Quarrying} &
      {\color[HTML]{000000} 0.25} &
      {\color[HTML]{000000} 0.50} &
      {\color[HTML]{000000} 0.75} &
      {\color[HTML]{000000} 1.00} \\
    \rowcolor[HTML]{F2F2F2} 
    {\color[HTML]{000000} Manufacturing} &
      {\color[HTML]{000000} 0.25} &
      {\color[HTML]{000000} 0.50} &
      {\color[HTML]{000000} 0.75} &
      {\color[HTML]{000000} 1.00} \\
    {\color[HTML]{000000} Electricity, Gas, Water Supply \& Other Utility} &
      {\color[HTML]{000000} 0.50} &
      {\color[HTML]{000000} 0.75} &
      {\color[HTML]{000000} 1.00} &
      {\color[HTML]{000000} 1.09} \\
    \rowcolor[HTML]{F2F2F2} 
    {\color[HTML]{000000} Construction} &
      {\color[HTML]{000000} 0.25} &
      {\color[HTML]{000000} 0.50} &
      {\color[HTML]{000000} 0.75} &
      {\color[HTML]{000000} 1.00} \\
    {\color[HTML]{000000} Trade, Hotels, Transport, Communication and Broadcasting} &
      {\color[HTML]{000000} 0.25} &
      {\color[HTML]{000000} 0.50} &
      {\color[HTML]{000000} 0.75} &
      {\color[HTML]{000000} 1.00} \\
    \rowcolor[HTML]{F2F2F2} 
    {\color[HTML]{000000} Financial, Real Estate and Professional Services} &
      {\color[HTML]{000000} 0.50} &
      {\color[HTML]{000000} 0.75} &
      {\color[HTML]{000000} 1.00} &
      {\color[HTML]{000000} 1.07} \\
    {\color[HTML]{000000} Public Administration, Defence and Other Services} &
      {\color[HTML]{000000} 0.80} &
      {\color[HTML]{000000} 1.00} &
      {\color[HTML]{000000} 1.00} &
      {\color[HTML]{000000} 1.00} \\ \bottomrule
    \end{tabular}
    \end{table}

\begin{table}[]
  \centering
  \caption{Scenario 1 - Quarterly GVA in Major Sectors of Indian Economy during 2020-21 in Lakh Crore INR. (Author's Estimation) }
  \label{tab:S1QGVA202021}
  \resizebox{\textwidth}{!}{%
  \begin{tabular}{@{}lrrrrr@{}}
  \toprule
  {\color[HTML]{000000} Quarter} &
    {\color[HTML]{000000} \textbf{Q1}} &
    {\color[HTML]{000000} \textbf{Q2}} &
    {\color[HTML]{000000} \textbf{Q3}} &
    {\color[HTML]{000000} \textbf{Q4}} &
    {\color[HTML]{000000} \textbf{Total}} \\ \midrule
  \rowcolor[HTML]{F2F2F2} 
  {\color[HTML]{000000} Agriculture, Forestry and Fishing} &
    {\color[HTML]{000000} 329436.00} &
    {\color[HTML]{000000} 274821.75} &
    {\color[HTML]{000000} 609105.00} &
    {\color[HTML]{000000} 500676.82} &
    {\color[HTML]{000000} 1714039.57} \\
  {\color[HTML]{000000} Mining \& Quarrying} &
    {\color[HTML]{000000} 23194.25} &
    {\color[HTML]{000000} 34945.00} &
    {\color[HTML]{000000} 85429.00} &
    {\color[HTML]{000000} 112182.74} &
    {\color[HTML]{000000} 255750.99} \\
  \rowcolor[HTML]{F2F2F2} 
  {\color[HTML]{000000} Manufacturing} &
    {\color[HTML]{000000} 143602.75} &
    {\color[HTML]{000000} 288592.00} &
    {\color[HTML]{000000} 559335.00} &
    {\color[HTML]{000000} 592498.43} &
    {\color[HTML]{000000} 1584028.18} \\
  {\color[HTML]{000000} Electricity, Gas, Water Supply \& Other Utility} &
    {\color[HTML]{000000} 40814.00} &
    {\color[HTML]{000000} 59643.75} &
    {\color[HTML]{000000} 72817.00} &
    {\color[HTML]{000000} 70776.31} &
    {\color[HTML]{000000} 244051.06} \\
  \rowcolor[HTML]{F2F2F2} 
  {\color[HTML]{000000} Construction} &
    {\color[HTML]{000000} 65913.25} &
    {\color[HTML]{000000} 122431.50} &
    {\color[HTML]{000000} 260170.00} &
    {\color[HTML]{000000} 256467.50} &
    {\color[HTML]{000000} 704982.25} \\
  {\color[HTML]{000000} Trade, Hotels, Transport, Communication and Broadcasting} &
    {\color[HTML]{000000} 161056.00} &
    {\color[HTML]{000000} 310576.50} &
    {\color[HTML]{000000} 484109.25} &
    {\color[HTML]{000000} 635517.08} &
    {\color[HTML]{000000} 1591258.83} \\
  \rowcolor[HTML]{F2F2F2} 
  {\color[HTML]{000000} Financial, Real Estate and Professional Services} &
    {\color[HTML]{000000} 405250.00} &
    {\color[HTML]{000000} 655359.75} &
    {\color[HTML]{000000} 643389.00} &
    {\color[HTML]{000000} 628206.59} &
    {\color[HTML]{000000} 2332205.34} \\
  {\color[HTML]{000000} Public Administration, Defence and Other Services} &
    {\color[HTML]{000000} 336952.80} &
    {\color[HTML]{000000} 461487.00} &
    {\color[HTML]{000000} 474947.00} &
    {\color[HTML]{000000} 433358.80} &
    {\color[HTML]{000000} 1706745.60} \\
  \rowcolor[HTML]{F2F2F2} 
  {\color[HTML]{000000} Total Gross Value Added at Basic Price} &
    {\color[HTML]{000000} 1506219.05} &
    {\color[HTML]{000000} 2207857.25} &
    {\color[HTML]{000000} 3189301.25} &
    {\color[HTML]{000000} 3229684.28} &
    {\color[HTML]{000000} 10133061.83} \\ \bottomrule
  \end{tabular}
  }
  \end{table}
 
  \begin{table}[]
    \centering
  \caption{Scenario 1 - Growth of GVA in 2020-21 (Author's Estimation)}
  \label{tab:S1QG202021}
  \begin{tabular}{@{}lrrrrr@{}}
    \toprule
    {\color[HTML]{000000} Quarter} &
      {\color[HTML]{000000} \textbf{Q1}} &
      {\color[HTML]{000000} \textbf{Q2}} &
      {\color[HTML]{000000} \textbf{Q3}} &
      {\color[HTML]{000000} \textbf{Q4}} &
      {\color[HTML]{000000} \textbf{Total}} \\ \midrule
    \rowcolor[HTML]{F2F2F2} 
    {\color[HTML]{000000} Agriculture, Forestry and Fishing} &
      {\color[HTML]{000000} -25\%} &
      {\color[HTML]{000000} -25\%} &
      {\color[HTML]{000000} 0\%} &
      {\color[HTML]{000000} 3\%} &
      {\color[HTML]{000000} -10\%} \\
    {\color[HTML]{000000} Mining \& Quarrying} &
      {\color[HTML]{000000} -75\%} &
      {\color[HTML]{000000} -50\%} &
      {\color[HTML]{000000} 0\%} &
      {\color[HTML]{000000} 5\%} &
      {\color[HTML]{000000} -28\%} \\
    \rowcolor[HTML]{F2F2F2} 
    {\color[HTML]{000000} Manufacturing} &
      {\color[HTML]{000000} -75\%} &
      {\color[HTML]{000000} -50\%} &
      {\color[HTML]{000000} 0\%} &
      {\color[HTML]{000000} 2\%} &
      {\color[HTML]{000000} -31\%} \\
    {\color[HTML]{000000} Electricity, Gas, Water Supply \& Other Utility} &
      {\color[HTML]{000000} -50\%} &
      {\color[HTML]{000000} -25\%} &
      {\color[HTML]{000000} 0\%} &
      {\color[HTML]{000000} 9\%} &
      {\color[HTML]{000000} -18\%} \\
    \rowcolor[HTML]{F2F2F2} 
    {\color[HTML]{000000} Construction} &
      {\color[HTML]{000000} -75\%} &
      {\color[HTML]{000000} -50\%} &
      {\color[HTML]{000000} 0\%} &
      {\color[HTML]{000000} 5\%} &
      {\color[HTML]{000000} -30\%} \\
    {\color[HTML]{000000} Trade, Hotels, Transport, Communication and Broadcasting} &
      {\color[HTML]{000000} -75\%} &
      {\color[HTML]{000000} -50\%} &
      {\color[HTML]{000000} -25\%} &
      {\color[HTML]{000000} 0\%} &
      {\color[HTML]{000000} -38\%} \\
    \rowcolor[HTML]{F2F2F2} 
    {\color[HTML]{000000} Financial, Real Estate and Professional Services} &
      {\color[HTML]{000000} -50\%} &
      {\color[HTML]{000000} -25\%} &
      {\color[HTML]{000000} 0\%} &
      {\color[HTML]{000000} 7\%} &
      {\color[HTML]{000000} -20\%} \\
    {\color[HTML]{000000} Public Administration, Defence and Other Services} &
      {\color[HTML]{000000} -20\%} &
      {\color[HTML]{000000} 0\%} &
      {\color[HTML]{000000} 0\%} &
      {\color[HTML]{000000} 0\%} &
      {\color[HTML]{000000} -5\%} \\
    \rowcolor[HTML]{F2F2F2} 
    {\color[HTML]{000000} Total Gross Value Added at Basic Price} &
      {\color[HTML]{000000} -55\%} &
      {\color[HTML]{000000} -33\%} &
      {\color[HTML]{000000} -5\%} &
      {\color[HTML]{000000} 3\%} &
      {\color[HTML]{000000} -23\%} \\ \bottomrule
    \end{tabular}
    \end{table}

\begin{table}[]
  \centering
  \caption{Scenario 2 - Quarterly GVA in Major Sectors of Indian Economy during 2020-21 in Lakh Crore INR. (Author's Estimation)}
  \label{tab:S2QGVA202021}
  \resizebox{\textwidth}{!}{%
  \begin{tabular}{@{}lrrrrr@{}}
  \toprule
  {\color[HTML]{000000} Quarter} &
    {\color[HTML]{000000} \textbf{Q1}} &
    {\color[HTML]{000000} \textbf{Q2}} &
    {\color[HTML]{000000} \textbf{Q3}} &
    {\color[HTML]{000000} \textbf{Q4}} &
    {\color[HTML]{000000} \textbf{Total}} \\ \midrule
  \rowcolor[HTML]{F2F2F2} 
  {\color[HTML]{000000} Agriculture, Forestry and Fishing} &
    {\color[HTML]{000000} 329436} &
    {\color[HTML]{000000} 274821.8} &
    {\color[HTML]{000000} 609105} &
    {\color[HTML]{000000} 500676.8} &
    {\color[HTML]{000000} 1714040} \\
  {\color[HTML]{000000} Mining \& Quarrying} &
    {\color[HTML]{000000} 23194.25} &
    {\color[HTML]{000000} 34945} &
    {\color[HTML]{000000} 64071.75} &
    {\color[HTML]{000000} 107146.8} &
    {\color[HTML]{000000} 229357.8} \\
  \rowcolor[HTML]{F2F2F2} 
  {\color[HTML]{000000} Manufacturing} &
    {\color[HTML]{000000} 143602.8} &
    {\color[HTML]{000000} 288592} &
    {\color[HTML]{000000} 419501.3} &
    {\color[HTML]{000000} 579744.1} &
    {\color[HTML]{000000} 1431440} \\
  {\color[HTML]{000000} Electricity, Gas, Water Supply \& Other Utility} &
    {\color[HTML]{000000} 40814} &
    {\color[HTML]{000000} 59643.75} &
    {\color[HTML]{000000} 72817} &
    {\color[HTML]{000000} 70776.31} &
    {\color[HTML]{000000} 244051.1} \\
  \rowcolor[HTML]{F2F2F2} 
  {\color[HTML]{000000} Construction} &
    {\color[HTML]{000000} 65913.25} &
    {\color[HTML]{000000} 122431.5} &
    {\color[HTML]{000000} 195127.5} &
    {\color[HTML]{000000} 243097.2} &
    {\color[HTML]{000000} 626569.4} \\
  {\color[HTML]{000000} Trade, Hotels, Transport, Communication and Broadcasting} &
    {\color[HTML]{000000} 161056} &
    {\color[HTML]{000000} 310576.5} &
    {\color[HTML]{000000} 484109.3} &
    {\color[HTML]{000000} 635517.1} &
    {\color[HTML]{000000} 1591259} \\
  \rowcolor[HTML]{F2F2F2} 
  {\color[HTML]{000000} Financial, Real Estate and Professional Services} &
    {\color[HTML]{000000} 405250} &
    {\color[HTML]{000000} 655359.8} &
    {\color[HTML]{000000} 643389} &
    {\color[HTML]{000000} 628206.6} &
    {\color[HTML]{000000} 2332205} \\
  {\color[HTML]{000000} Public Administration, Defence and Other Services} &
    {\color[HTML]{000000} 336952.8} &
    {\color[HTML]{000000} 461487} &
    {\color[HTML]{000000} 474947} &
    {\color[HTML]{000000} 433358.8} &
    {\color[HTML]{000000} 1706746} \\
  \rowcolor[HTML]{F2F2F2} 
  {\color[HTML]{000000} Total Gross Value Added at Basic Price} &
    {\color[HTML]{000000} 1506219} &
    {\color[HTML]{000000} 2207857} &
    {\color[HTML]{000000} 2963068} &
    {\color[HTML]{000000} 3198524} &
    {\color[HTML]{000000} 9875668} \\ \bottomrule
  \end{tabular}
  }
  \end{table}

\begin{table}[]
  \centering
\caption{Scenario 2 - Growth of GVA in 2020-21 (Author's Estimation)}
\label{tab:S2QG202021}
    \begin{tabular}{@{}llllll@{}}
  \toprule
  {\color[HTML]{000000} Quarter} &
    {\color[HTML]{000000} \textbf{Q1}} &
    {\color[HTML]{000000} \textbf{Q2}} &
    {\color[HTML]{000000} \textbf{Q3}} &
    {\color[HTML]{000000} \textbf{Q4}} &
    {\color[HTML]{000000} \textbf{Total}} \\ \midrule
  \rowcolor[HTML]{F2F2F2} 
  {\color[HTML]{000000} Agriculture, Forestry and Fishing} &
    {\color[HTML]{000000} -25\%} &
    {\color[HTML]{000000} -25\%} &
    {\color[HTML]{000000} 0\%} &
    {\color[HTML]{000000} 3\%} &
    {\color[HTML]{000000} -10\%} \\
  {\color[HTML]{000000} Mining \& Quarrying} &
    {\color[HTML]{000000} -75\%} &
    {\color[HTML]{000000} -50\%} &
    {\color[HTML]{000000} -25\%} &
    {\color[HTML]{000000} 0\%} &
    {\color[HTML]{000000} -35\%} \\
  \rowcolor[HTML]{F2F2F2} 
  {\color[HTML]{000000} Manufacturing} &
    {\color[HTML]{000000} -75\%} &
    {\color[HTML]{000000} -50\%} &
    {\color[HTML]{000000} -25\%} &
    {\color[HTML]{000000} 0\%} &
    {\color[HTML]{000000} -38\%} \\
  {\color[HTML]{000000} Electricity, Gas, Water Supply \& Other Utility} &
    {\color[HTML]{000000} -50\%} &
    {\color[HTML]{000000} -25\%} &
    {\color[HTML]{000000} 0\%} &
    {\color[HTML]{000000} 9\%} &
    {\color[HTML]{000000} -18\%} \\
  \rowcolor[HTML]{F2F2F2} 
  {\color[HTML]{000000} Construction} &
    {\color[HTML]{000000} -75\%} &
    {\color[HTML]{000000} -50\%} &
    {\color[HTML]{000000} -25\%} &
    {\color[HTML]{000000} 0\%} &
    {\color[HTML]{000000} -38\%} \\
  {\color[HTML]{000000} Trade, Hotels, Transport, Communication and Broadcasting} &
    {\color[HTML]{000000} -75\%} &
    {\color[HTML]{000000} -50\%} &
    {\color[HTML]{000000} -25\%} &
    {\color[HTML]{000000} 0\%} &
    {\color[HTML]{000000} -38\%} \\
  \rowcolor[HTML]{F2F2F2} 
  {\color[HTML]{000000} Financial, Real Estate and Professional Services} &
    {\color[HTML]{000000} -50\%} &
    {\color[HTML]{000000} -25\%} &
    {\color[HTML]{000000} 0\%} &
    {\color[HTML]{000000} 7\%} &
    {\color[HTML]{000000} -20\%} \\
  {\color[HTML]{000000} Public Administration, Defence and Other Services} &
    {\color[HTML]{000000} -20\%} &
    {\color[HTML]{000000} 0\%} &
    {\color[HTML]{000000} 0\%} &
    {\color[HTML]{000000} 0\%} &
    {\color[HTML]{000000} -5\%} \\
  \rowcolor[HTML]{F2F2F2} 
  {\color[HTML]{000000} Total Gross Value Added at Basic Price} &
    {\color[HTML]{000000} -55\%} &
    {\color[HTML]{000000} -33\%} &
    {\color[HTML]{000000} -12\%} &
    {\color[HTML]{000000} 2\%} &
    {\color[HTML]{000000} -25\%} \\ \bottomrule
  \end{tabular}
  \end{table}

  \clearpage
  \bibliographystyle{unsrt}
  \bibliography{paper}

\begin{thebibliography}{1}

\bibitem{IMF:2020}
IMF.
\newblock {\em World Economic Outlook, April 2020: The Great Lockdown}, 2020
  (accessed May 10, 2020).

\bibitem{ET1:2020}
The~Economic Times.
\newblock Fitch ratings sees india growth slipping to 0.8\% in fy21, 2020
  (accessed May 10, 2020).

\bibitem{TheHindu:2020}
C.~Rangarajan and D.~K. Srivastav.
\newblock Slower growth and a tighter fiscal, 2020 (accessed May 10, 2020).

\bibitem{ET4:2020}
The~Economic Times.
\newblock Adb expects india gdp to slip to 4 pc in 2020-21; projects strong
  recovery next fiscal, 2020 (accessed May 10, 2020).

\bibitem{ET2:2020}
The~Economic Times.
\newblock Lockdown pulls down power consumption by 22.75\% to 85.05 bu in
  april, 2020 (accessed May 10, 2020).

\bibitem{ET3:2020}
The~Economic Times.
\newblock Core sector output shrinks 6.5
  on the economy, 2020 (accessed May 10, 2020).

\bibitem{CMIE:2020}
CMIE.
\newblock Net direct tax collection up 36.5\% in april 2020, 2020 (accessed May
  10, 2020).

\bibitem{BusinessToday:2020}
Nirbhay Kumar.
\newblock Precipitous 80-90\% fall in states' april gst collections; centre may
  see massive drop, 2020 (accessed May 10, 2020).

\end{thebibliography}



\end{document}